\documentclass{article}
\usepackage{spconf,amsmath,graphicx}
\usepackage{color}
\usepackage{multirow}
\usepackage{xcolor}
\usepackage{multicol}
\usepackage{booktabs}
\usepackage{graphicx}
\usepackage{enumitem}
\usepackage{caption}
\usepackage{subcaption}

\usepackage{lipsum}

\let\OLDthebibliography\thebibliography
\renewcommand\thebibliography[1]{
  \OLDthebibliography{#1}
  \setlength{\parskip}{0pt}
  \setlength{\itemsep}{0pt plus 0.1ex}
}

\usepackage{hyperref}
\hypersetup{
    colorlinks=true,
    urlcolor=magenta,
}

\def\x{$\times$}

\definecolor{fastcolor}{RGB}{121,178,128}

\definecolor{blocks_color}{RGB}{186,115,115}
\definecolor{blockt_color}{RGB}{88,138,122}
\newcommand{\blockscolor}[1]{\textcolor{blocks_color}{#1}}
\newcommand{\blocktcolor}[1]{\textcolor{blockt_color}{#1}}

\newcommand{\outsizes}[5]{\multirow{#5}{*}{\(\begin{array}{c} \text{\emph{Slow}}: \text{#1$\times$#2}\\[-.1em] \text{\emph{Fast}}: \text{#3$\times$#4}\end{array}\)}}
\newcommand{\blocks}[3]{\multirow{3}{*}{\(\left[\begin{array}{c}\text{\blockscolor{1$\times$1}, #2}\\[-.1em] \text{\blockscolor{1$\times$3}, #2}\\[-.1em] \text{\blockscolor{1$\times$1}, #1}\end{array}\right]\)$\times$#3}}
\newcommand{\blockt}[3]{\multirow{3}{*}{\(\left[\begin{array}{c}\text{\blocktcolor{3$\times$1}, #2}\\[-.1em] \text{\blocktcolor{1$\times$3}, #2}\\[-.1em] \text{\blocktcolor{1$\times$1}, #1}\end{array}\right]\)$\times$#3}}

\newlength\savewidth\newcommand\shline{\noalign{\global\savewidth\arrayrulewidth
		\global\arrayrulewidth 1pt}\hline\noalign{\global\arrayrulewidth\savewidth}}

\title{SLOW-FAST AUDITORY STREAMS FOR AUDIO RECOGNITION}

\name{Evangelos Kazakos$^{\star}$ \qquad Arsha Nagrani$^{\dagger\ddag}$ \qquad Andrew Zisserman$^{\dagger}$ \qquad Dima Damen$^{\star}$}
\address{$^{\star}$ Department of Computer Science, University of Bristol\\
      $^{\dagger}$Visual Geometry Group, University of Oxford
      \thanks{\hspace{-10pt}$\ddag$ Now at Google Research.}}

\begin{document}

\maketitle

\begin{abstract}
We propose a two-stream convolutional network for audio recognition, that operates on time-frequency spectrogram inputs. 
Following similar success in visual recognition, we learn Slow-Fast auditory streams with separable convolutions and multi-level lateral connections. 
The Slow pathway has high channel capacity while the Fast pathway operates at a fine-grained temporal resolution.
We showcase the importance of our two-stream proposal on two diverse datasets: VGG-Sound and EPIC-KITCHENS-100, and achieve state-of-the-art results on both.
\end{abstract}


\section{Introduction}
\label{sec:intro}

Recognising objects, interactions and activities from audio is distinct from prior efforts for scene audio recognition, due to the need for recognising sound-emitting objects (e.g.\ alarm clock, coffee-machine), sounds generated from interactions with objects (e.g.\ put down a glass, close drawer), and activities (e.g.\ wash, fry). 
This introduces challenges related to variable-length audio associated with these activities. Some can be momentary (e.g.\ close) while others are repetitive over a longer period (e.g.\ fry), and many exhibit intra-class variations (e.g.\ cut onion vs cut cheese). Background or irrelevant sounds are often captured with these activities.
We focus on two activity-based datasets, VGG-Sound \cite{vggsound} and EPIC-KITCHENS \cite{Damen2020RESCALING}, captured from YouTube and egocentric videos respectively, and target activity recognition solely from the audio signal associated with these videos.

There is strong evidence in neuroscience for the existence of two streams in the human auditory system, the ventral stream for identifying sound-emitting objects and the dorsal streams for locating these objects. Studies \cite{santoro,zulfiqar} suggest the ventral stream accordingly exhibits high spectral resolution for object identification, while the dorsal stream has a high temporal resolution and operates at a higher sampling rate. 

Using this evidence as the driving force for designing our architecture, and inspired by a similar vision-based architecture~\cite{Feichtenhofer_2019_ICCV}, we propose two streams for auditory recognition: a Slow and a Fast stream, that realise some of the properties of the ventral and dorsal auditory pathways respectively. 
Our streams are variants of residual networks and use 2D separable convolutions that operate on frequency and time independently. 
The streams are fused in multiple representation levels with lateral connections from the Fast to the Slow stream, and the final representation is obtained by concatenating the global average pooled representations for action recognition. 

The contributions of this paper are the following: i) we propose a novel two-stream architecture for auditory recognition that respects evidence in neuroscience; ii) we achieve state-of-the-art results on both EPIC-KITCHENS and  VGG-Sound; and finally iii) we showcase the importance of fusing our specialised streams through an ablation analysis. Our pretrained models and code is available at \url{https://github.com/ekazakos/auditory-slow-fast}.

\section{Related work}
\label{sec:related}

\textbf{Single-stream architectures}. A common approach in audio recognition for both scene and activity recognition, is to use a single-stream convolutional architecture~\cite{Salamon2017,Hershey2017, soundnet}. 
SoundNet~\cite{soundnet} uses 1D ConvNet trained in a teacher-student manner, and fine-tuned for acoustic scene classification. Single-stream 2D ConvNets have been extensively used by high-ranked entries of DCASE challenges~\cite{Suh2020,Hu2020,KoutiniEDW19,Chen2019,Dorfer2018,Liping2018}, for acoustic scene classification.
These consider spectograms as input and utilise 2D convolutions with square $k\times k$ filters, processing frequency and time together \cite{Salamon2017,Hershey2017, Suh2020,Hu2020,KoutiniEDW19,Chen2019,Dorfer2018,Liping2018}, similarly to image ConvNets. However, symmetric filtering in frequency and time might not be optimal as the statistics of spectrograms are not homogeneous. 
One alternative is to utilise rectangular $k \times m$ filters as in \cite{pons2017,Huzaifah17}.
Another is separable convolutions with $1\times k$ and $k\times 1$ filters, which have recently been used in audio~\cite{xiao2020audiovisual,dap}. 

\begin{figure*}[t]
    \centering
    \includegraphics[width=18cm]{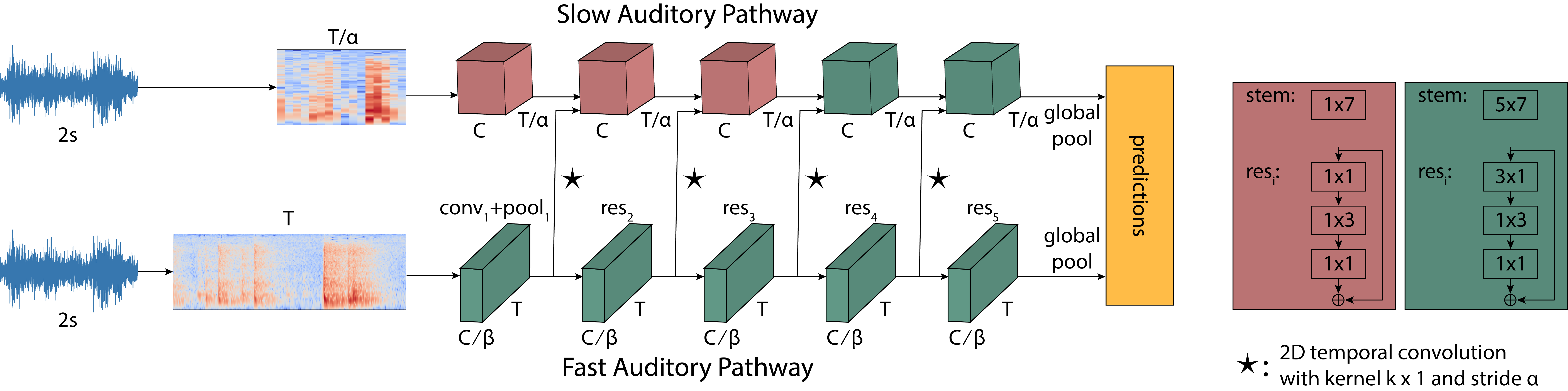}
    \vspace*{-12pt}
    \caption{Proposed Slow-Fast architecture. Strided input (by $\alpha$) to the Slow pathway, along with increased channels. The Fast pathway has less channels (by $\beta$). Right: two types of residual blocks with separable convolutions (brown vs green).}
\label{fig:arch}
\end{figure*}

\noindent\textbf{Multi-stream architectures}. Late fusion of multiple streams for audio recognition was used in~\cite{SuZWM19,Li2019,Bhatt2018,Wang2020_t1,mcdonnell_2020,Wu2020,Han2019}. Most approaches utilise modality-specific streams~\cite{SuZWM19,Li2019,Bhatt2018,Wang2020_t1}. 
In addition to late fusion,~\cite{Li2019,Bhatt2018} integrate multi-level fusion in their architecture in the form of attention. 

In \cite{mcdonnell_2020,Wu2020,Han2019}, all streams digest the same input.
In~\cite{mcdonnell_2020}, one stream takes as input low frequencies and the second inputs high frequencies. 
\cite{Wu2020} applies median filtering with different kernels at the input of each stream to model long duration sound events, medium, and short duration impulses separately. In~\cite{Han2019} , 1D convolutions are used with different dilation rates at each stream to model convolutional streams that operate on different temporal resolutions.
The architectures of these multiple streams remain identical.

Similar to these works, we propose to utilise two-streams that consider the same input. Different from these, we design each stream with varying number of channels and temporal resolution, in addition to convolutional separation. Furthermore, we integrate the streams through multi-level fusion.

\section{Network Architecture}
\label{sec:audioslowfast}

Next, we describe in detail the design principles of our architecture, depicted in Figure \ref{fig:arch}.
The Slow stream operates on a low sampling rate with high channel capacity to capture frequency semantics, while the Fast stream operates on a high sampling rate with more temporal convolutions and less channels to capture temporal patterns. 

\noindent \textbf{Input}. Both streams operate on the same audio length, from which a log-mel-spectrogram is extracted. The Fast stream takes as input the whole log-mel-spectrogram without any striding, while the Slow stream uses a temporal stride of $\alpha$ on the input log-mel-spectrogram, where $a\geq1$. 

\begin{table}[t]
	\scriptsize
	\centering
	\resizebox{\columnwidth}{!}{
		\begin{tabular}{c|c|c|c}
			stage & \emph{Slow} pathway &  \emph{Fast} pathway & output sizes $T$\x$F$ \\
			\shline
			\multirow{1}{*}{spectrogram} & - & - & 400\x128\\
			\hline
			\multirow{2}{*}{data layer} & \multirow{2}{*}{stride 4, 1} & \multirow{2}{*}{stride 1, 1} &  \outsizes{{100}}{128}{400}{128}{2}   \\
			&  &  \\
			\hline
			\multirow{2}{*}{conv$_1$} & \multicolumn{1}{c|}{\blockscolor{1\x7}, {64}} & \multicolumn{1}{c|}{\blocktcolor{5\x7}, 8} &  \outsizes{{50}}{64}{200}{64}{2}   \\
			& stride 2, 2 & stride 2, 2  \\
			\hline
			\multirow{2}{*}{pool$_1$}  & \multicolumn{1}{c|}{3\x3 max} & \multicolumn{1}{c|}{3\x3 max} &  \outsizes{{25}}{32}{100}{32}{2} \\
			& stride 2, 2 & stride 2, 2 & \\
			\hline
			\multirow{3}{*}{res$_2$} & \blocks{{256}}{{64}}{3} & \blockt{32}{8}{3} & \outsizes{{25}}{32}{100}{32}{3}  \\
			&  & \\
			&  & \\
			\hline
			\multirow{3}{*}{res$_3$} & \blocks{{512}}{{128}}{4} &  \blockt{64}{16}{4}  & \outsizes{{25}}{16}{100}{16}{3}  \\
			&  & \\
			&  & \\
			\hline
			\multirow{3}{*}{res$_4$} & \blockt{{1024}}{{256}}{6} & \blockt{128}{32}{6} &  \outsizes{{25}}{8}{100}{8}{3}  \\
			&  & \\
			&  & \\
			\hline
			\multirow{3}{*}{res$_5$} & \blockt{{2048}}{{512}}{3} & \blockt{256}{64}{3} &   \outsizes{{25}}{4}{100}{4}{3} \\
			&  & \\
			&  & \\
			\hline
			\multicolumn{3}{c|}{global average pool, concatenate, fc}  & \# classes \\
	\end{tabular}}
	\caption{Architecture details for Fig.~\ref{fig:arch}}
	\label{tab:arch}
		\vspace{-12pt}
\end{table}

\noindent\textbf{Slow and Fast streams}. The two streams are variants of ResNet50 \cite{He_2016_CVPR}. Each stream is comprised of an initial convolutional block with a pooling layer followed by 4 residual stages, where each stage contains multiple residual blocks. 
The two streams differ in their ability to capture frequency semantics and temporal patterns.
The details of each stream including the number of blocks per stage and numbers of channels can be seen in Table~\ref{tab:arch}. 

The Slow stream has a high channel capacity, with $\beta$ times more channels than the Fast stream, while operating on a low sampling rate. As the input spectrogram is strided temporally by $\alpha$, the intermediate feature maps have a lower temporal resolution. Moreover, the Slow stream has temporal convolutions only in $\text{res}_{\text{4}}$ and $\text{res}_{\text{5}}$ (see the brown and green blocks in Fig.~\ref{fig:arch} right). By restricting the temporal resolution and the temporal kernels of the Slow stream while keeping a high channel capacity, this stream can focus on learning frequency semantics. 

The Fast stream on the other hand uses no temporal striding in the input. Therefore, the intermediate feature maps have a higher temporal resolution, with temporal convolutions throughout the stream. With a high temporal resolution and more temporal kernels while having less channels, it is easier for the Fast stream to focus on learning temporal patterns.

\noindent\textbf{Separable convolutions}. 
We use separable convolutions in frequency and time as can be seen in the green block in Fig.~\ref{fig:arch} right. We break a $3\times3$ kernel in two kernels, $3\times1$ followed by $1\times3$. 
Separable convolutions have proven useful for video recognition \cite{Tran_2018_CVPR}. We utilise them with the motivation to separately attend to time and frequency of the input signal. We contrast separable convolutions to two-dimensional filters that convolve across both frequency and time. 

\noindent\textbf{Multi-level fusion}. 
Following the approach in~\cite{Feichtenhofer_2019_ICCV}, we fuse the information from the Fast to the Slow stream with lateral connections, at multiple levels. 
We first apply a 2D temporal convolution with a kernel $7\times1$ and a stride~of~$\alpha$ to the output of the Fast stream to match the Slow stream sampling rate, and then we concatenate the downsampled feature map with the Slow stream feature map. Fusion is applied after $\text{pool}_\text{1}$ and each residual stage.   

The final representation fed to the classifier is obtained by applying time-frequency global average pooling after the last convolutional layer of both Slow and Fast streams and concantenating the pooled representations. We set $\alpha=4$ and $\beta=8$ in all our experiments.

\noindent\textbf{Differences compared to visual Slow-Fast~\cite{Feichtenhofer_2019_ICCV}}.
Our two-stream architecture is inspired by its visual counterpart~\cite{Feichtenhofer_2019_ICCV} which produces state of the art results for visual action recognition. 
However, key differences are introduced:
Our input is 2D rather than 3D, as we operate on time-frequency while the visual Slow-Fast operates on time-space. Hence, we use 2D separable convolutions decomposed as $3\times1$ and $1\times3$ filters, whereas \cite{Feichtenhofer_2019_ICCV} uses 3D separable convolutions decomposed as $3\times1\times1$ and $1\times3\times3$ filters.
Additionally, the sampling rate for audio is  naturally significantly higher than that of video, e.g.\ 24kHz vs 50fps in EPIC-KITCHENS-100, and the dimensionality in video is significantly higher. Accordingly, the approach in~\cite{Feichtenhofer_2019_ICCV} only considers a few temporal samples (8 and 32 frames in the Slow and Fast streams respectively). In contrast, our audio spectogram (see Sec~\ref{subsec:protocol}) contains 100 and 400 temporal dimensions in the Slow and Fast streams respectively. To compensate for the high sampling rate of audio, we temporally downsample the representations of both streams by a factor of 4, using a temporal stride=2 in $\text{conv}_{\text{1}}$ and $\text{pool}_{\text{1}}$ of both streams. The remaining stages do not perform any temporal downsampling\footnote{In preliminary experiments, we tried different downsampling schemes, such as strided convolutions throughout the whole network but they resulted in inferior performance.}.

\begin{table*}[t!]
\centering
\resizebox{\textwidth}{!}{%
\begin{tabular}{@{}clrrrrrrrrrr@{}}
\toprule
                                           &                                          & \multicolumn{6}{c}{Overall}                           & \multicolumn{3}{c}{Unseen Participants} \\
                                                                                      \cmidrule(r){3-8}                                       \cmidrule(lr){9-11}                   
                                           &                                          & \multicolumn{3}{c}{Top-1 Accuracy (\%)} & \multicolumn{3}{c}{Top-5 Accuracy (\%)} & \multicolumn{3}{c}{Top-1 Accuracy (\%)}\\
                                                                                      \cmidrule(r){3-5}           \cmidrule(lr){6-8}          \cmidrule(lr){9-11}   
Split                                      & Model                                    & \multicolumn{1}{c}{Verb} & \multicolumn{1}{c}{Noun} & \multicolumn{1}{c}{Action} & \multicolumn{1}{c}{Verb} & \multicolumn{1}{c}{Noun} & \multicolumn{1}{c}{Action}& \multicolumn{1}{c}{Verb} & \multicolumn{1}{c}{Noun} & \multicolumn{1}{c}{Action} & \multicolumn{1}{r}{\# Param.}\\ \midrule
\multirow{7}{*}{\rotatebox{90}{\textbf{Val}}}

                                           & Damen et al. \cite{Damen2020RESCALING}                            																	&  42.63 & 22.35 & 14.48 & 75.84 & 44.60 & 28.23 & 35.40 & \textbf{16.34} & \textbf{9.20}   & 10.67M\\
                                           & Slow                                                                                              & 41.17 & 18.64 & 11.37 & 77.52 & 42.34 & 24.20 & 34.93 & 14.65 & 7.79  & 24.89M\\
                                           & Fast                                                                                              & 39.84 & 17.07 & 8.76 & 76.94 & 41.31 & 22.01 & 33.33 & 15.21 & 6.57 & 00.49M\\
                                           \cmidrule{2-12}
                                           & Two Slow Streams                                                                                             & 41.41 & 19.06 & 11.41 & 77.87 & 43.05 & 24.73 & 34.37 & 14.27 & 6.85 & 49.78M\\
                                            & Slow ResNet101                                                                                              & 42.24 & 19.35 & 12.12 & 78.14 & 42.83 & 25.30 & 37.37 & 13.90 & 7.61 & 46.11M\\
                                            \cmidrule{2-12}
                                           & Slow-Fast (late fusion) & 42.28 & 19.23 & 11.27 & 78.40 & 44.17 & 25.36 & 34.65 & 15.68 & 7.70 & 25.38M\\
            								& Slow-Fast (Proposed)                                                                                   & \textbf{46.05} & \textbf{22.95} & \textbf{15.22} & \textbf{80.01} & \textbf{47.98} & \textbf{30.24} & \textbf{37.56} & \textbf{16.34} & 8.83 & 26.88M\\
                                            \cmidrule{1-12} \morecmidrules \cmidrule{1-12}
\multirow{2}{*}{\rotatebox{90}{\textbf{Test}}}

                                           & Damen et al. \cite{Damen2020RESCALING}																								& 42.12 & 21.51 & 14.76 & 75.06 & 41.12 & 25.86 & 37.45 & 17.74 & 11.63 & 10.67M\\
                                           & Slow-Fast (Proposed)                                                                                   & \textbf{46.47} & \textbf{22.77} & \textbf{15.44} & \textbf{78.30} & \textbf{44.91} & \textbf{28.56} & \textbf{42.48} & \textbf{20.12} & \textbf{12.92} & 26.88M\\
                                          \bottomrule
\end{tabular}}
\caption{Results on EPIC-KITCHENS-100. We provide an ablation study over the Val set, as well as report results on the Test set showing improvement over the published state-of-the-art in audio recognition. \# Parameters per model is also shown.}
\vspace*{-6pt}
\label{tab:epic_res}
\end{table*}

\begin{table}[t!]
\centering
\resizebox{\linewidth}{!}{%
\begin{tabular}{@{}lrrrrr@{}}
\toprule
                                     Model                                     & \multicolumn{1}{c}{Top-1} & \multicolumn{1}{c}{Top-5} & \multicolumn{1}{c}{mAP}           & \multicolumn{1}{c}{AUC}    & \multicolumn{1}{r}{d-prime}\\ \midrule
                                           Chen et al. \cite{vggsound}     & 51.00 & 76.40 & 0.532 & 0.973 & 2.735\\
                                           McDonnell \& Gao \cite{mcdonnell_2020} & 39.74 & 71.65 & 0.403 & 0.963 & 2.532\\
                                           Slow      & 45.20 & 72.53 & 0.472 & 0.967 & 2.607\\
                                                                                     Fast     & 41.44 & 70.68 & 0.442 & 0.966 & 2.576\\ \hline
                                           Two Slow Streams     & 45.80 & 72.78 & 0.482 & 0.969 & 2.633\\
                                         Slow ResNet101       & 45.60 & 72.27 & 0.476 & 0.968 & 2.615\\ \hline
            							 Slow-Fast (late fusion)	& 46.75 & 73.90 & 0.498 & 0.971 & 2.671\\
            							 Slow-Fast (Proposed)     & \textbf{52.46} & \textbf{78.12} & \textbf{0.544} & \textbf{0.974} & \textbf{2.761}\\

                                          \bottomrule
\end{tabular}}
\caption{Results on VGG-Sound. We compare to published results and show ablations.}
\vspace*{-12pt}
\label{tab:vggsound_res}
\end{table}

\section{Experiments}
\label{sec:experiments}
 
\subsection{Datasets}
\label{subsec:datasets}

\textbf{VGG-Sound}. VGG-Sound \cite{vggsound} is a large-scale audio dataset obtained from YouTube. It contains over 200k clips of 10s for 309 classes  
capturing human actions, sound-emitting objects as well as interactions. These are visually-grounded where sound emitting objects are visible in the corresponding video clip, utilising image classifiers to find correspondence between sound and image labels. Audio is sampled at 16kHz.

\noindent\textbf{EPIC-KITCHENS-100}. EPIC-KITCHENS-100~\cite{Damen2020RESCALING} is the largest egocentric audio-visual dataset, containing unscripted daily activities in kitchen environments. The data are recorded in 45 different kitchens. It contains 100 hours of data, split across 700 untrimmed videos, and 90K trimmed action clips. These capture hand-object interactions as well as activities, formed as the combination of a verb and a noun (e.g.\ ``cut onion" and ``wash plate''), where there are 97 verb classes, 300 noun classes, and 4025 action classes (many verbs and nouns do not co-occur). The classes are highly unbalanced. Actions are mainly short-term (average action length is 2.6s with minimum length 0.25s). Audio is sampled at 24kHz.

\subsection{Experimental protocol}
\label{subsec:protocol}

\textbf{Feature extraction}.
We extract log-mel-spectrograms with 128 Mel bands using the Librosa library. For VGG-Sound, we use 5.12s of audio with a window of 20ms and a hop of 10ms, resulting in spectrograms of size $512\times 128$. For EPIC-KITCHENS-100, we use 2s of audio with a 10ms window and a 5ms hop, resulting in spectrograms of size $400\times 128$. For clips $<$ 2s in EPIC-KITCHENS-100, we duplicate the last time-frame of the log-mel-spectrogram.

\noindent\textbf{Train / Val details}. All models are trained using SGD with momentum set to 0.9 and cross-entropy loss. We train on EPIC-KITCHENS-100 as a multitask learning problem, as in~\cite{Damen2020RESCALING}, using two prediction heads, one for verbs and one for nouns. We train on VGG-Sound from random initialisation for 50 epochs and fine-tune on EPIC-KITCHENS-100 using the VGG-Sound pretrained models for 30 epochs. We drop the learning rate by 0.1 at epochs 30 and 40 for VGG-Sound, and at epochs 20 and 25 for EPIC-KITCHENS-100. For fine-tuning, we freeze Batch-Normalisation layers except the first one, as done in \cite{TSN2016ECCV}. For regularisation, we use dropout on the concatenation of Slow and Fast streams with probability 0.5, plus weight decay in all trainable layers using the value of $10^{-4}$. For data augmentation during training, we use the implementation of SpecAugment \cite{Park2019} from \cite{specaugment_repo} and set its parameters as follows: 2 frequency masks with F=27, 2 time masks with T=25, and time warp with W=5. During training we randomly extract one audio segment from each clip. During testing we average the predictions of 2 equally distanced segments for VGG-Sound, and 10 for EPIC-KITCHENS-100.

\noindent\textbf{Evaluation metrics}. For VGG-Sound, we follow the evaluation protocol of \cite{vggsound,Hershey2017} and report mAP, AUC, and d-prime, as defined in \cite{Hershey2017}. Additionally we report top-1/5\% accuracy. For EPIC-KITCHENS-100, we follow the evaluation protocol of~\cite{Damen2020RESCALING} and report top-1 and top-5 \% accuracy for the validation and test sets separately, as well as for the subset of unseen participants within val/test.

\noindent\textbf{Baselines and ablation study}. We compare to published state-of-the-art results in each dataset. For VGG-Sound, we also compare against~\cite{mcdonnell_2020} using their publicly available code, which is the closest work to ours in motivation, as it uses two audio streams separating input into low/high frequencies.

We also perform an ablation study investigating the importance of the two streams as follows:
\begin{itemize}[leftmargin=*,itemsep=-2ex,partopsep=1ex,parsep=2ex]
\vspace*{-6pt}
    \item Slow, Fast: We compare to each single stream individually.
    \item Enriched Slow stream: We combine two Slow streams with late fusion of predictions, as well as a deeper Slow stream (ResNet101 instead of ResNet50).
    \item Slow-Fast without multi-level fusion: Streams are fused by averaging their predictions, without lateral connections.
\end{itemize}

\subsection{Results}
\label{subsec:res}

\textbf{EPIC-KITCHENS-100} Our proposed network achieves state-of-the-art results as can be seen in Table~\ref{tab:epic_res} for both Val and Test. Our previous results~\cite{Damen2020RESCALING} use a TSN with BN-Inception architecture~\cite{TSN2016ECCV}, initialised from ImageNet, while here we utilise pre-training from VGG-Sound. Our proposed architecture outperforms \cite{Damen2020RESCALING} by a good margin. 
We report the ablation comparison using the published Val split.
The significant improvement in our proposed Slow-Fast architecture when compared to Slow and Fast streams independently shows that there is complementary information in the two streams that benefit audio recognition. The Slow stream performs better than Fast, due to the increased number of channels. When comparing to the enriched Slow architectures (see the last column of Table~\ref{tab:epic_res} for number of parameters), 
our proposed model still significantly outperforms these baselines, showcasing the need for the two different pathways. We conclude that the synergy of Slow and Fast streams is more important than simply increasing the number of parameters of the stronger Slow stream. Finally, our proposed architecture consistently outperforms late fusion, indicating the importance of multi-level fusion with lateral connections.

\noindent\textbf{VGG-Sound}. We report results in Table~\ref{tab:vggsound_res} comparing to state-of-the-art from~\cite{Chen2019}, which uses a single-stream ResNet50 architecture, \cite{mcdonnell_2020} which uses a ResNet variant with 19 layers as backbone for their two-stream architecture with significantly less parameters than our model at 3.2M parameters, as well as ablations of our model. We report the best performing model on the test set in each case. Our proposed Slow-Fast architecture outperforms \cite{vggsound} and \cite{mcdonnell_2020}. The rest of our observations on the ablations from EPIC-KITCHENS-100 hold for VGG-Sound as well, with a key difference: the gap in performance between single streams and our proposed two-stream architecture is even bigger for VGG-Sound, indicating more complementary information in the two streams. The fact that Slow-Fast outperforms Slow by such a large accuracy gap with an insignificant increase in parameters indicates the efficient interaction between Slow and Fast streams. 

\vspace*{-14pt}
\section{Conclusion}
\label{sec:conclusion}
\vspace*{-4pt}
We propose a two-stream architecture for audio recognition, inspired by the two pathways in the human auditory system, fusing Slow and Fast streams with multi-level lateral connections. We showcase the importance of our fusion architecture through ablations on two activity-based datasets, EPIC-KITCHENS-100 and VGG-Sound, achieving state-of-the-art performance. For future work, we will explore learning the stride parameter and assessing the impact of the number of channels. We hope that this work will pave the path for efficient multi-stream training in audio.

\noindent \textbf{Acknowledgements.} Publicly-available datasets were used for this work. Kazakos is supported by EPSRC DTP, Damen by EPSRC Fellowship UMPIRE (EP/T004991/1) and Nagrani by Google PhD fellowship. Research is also supported by Seebibyte (EP/M013774/1).

\bibliographystyle{IEEEbib}
\bibliography{strings,refs}

\begin{thebibliography}{10}

\bibitem{vggsound}
H.~{Chen}, W.~{Xie}, A.~{Vedaldi}, and A.~{Zisserman},
\newblock ``{VGGsound: A Large-Scale Audio-Visual Dataset},''
\newblock in {\em ICASSP}, 2020.

\bibitem{Damen2020RESCALING}
D.~Damen, H.~Doughty, G.~M. Farinella, , A.~Furnari, J.~Ma, E.~Kazakos,
  D.~Moltisanti, J.~Munro, T.~Perrett, W.~Price, and M~Wray,
\newblock ``Rescaling egocentric vision,''
\newblock {\em CoRR}, vol. abs/2006.13256, 2020.

\bibitem{santoro}
R.~Santoro, M.~Moerel, F.~D.~Martino, R.~Goebel, K.~Ugurbil, E.~Yacoub, and
  E.~Formisano,
\newblock ``Encoding of natural sounds at multiple spectral and temporal
  resolutions in the human auditory cortex,''
\newblock {\em PLOS Computational Biology}, vol. 10, no. 1, pp. 1--14, 2014.

\bibitem{zulfiqar}
I.~Zulfiqar, M.~Moerel, and E.~Formisano,
\newblock ``Spectro-temporal processing in a two-stream computational model of
  auditory cortex,''
\newblock {\em Frontiers in Computational Neuroscience}, vol. 13, pp. 95, 2020.

\bibitem{Feichtenhofer_2019_ICCV}
C.~Feichtenhofer, H.~Fan, J.~Malik, and K.~He,
\newblock ``Slowfast networks for video recognition,''
\newblock in {\em ICCV}, 2019.

\bibitem{Salamon2017}
J.~{Salamon} and J.~P. {Bello},
\newblock ``Deep convolutional neural networks and data augmentation for
  environmental sound classification,''
\newblock {\em IEEE Signal Processing Letters}, vol. 24, no. 3, pp. 279--283,
  2017.

\bibitem{Hershey2017}
S.~{Hershey}, S.~{Chaudhuri}, D.~P.~W. {Ellis}, J.~F. {Gemmeke}, A.~{Jansen},
  R.~C. {Moore}, M.~{Plakal}, D.~{Platt}, R.~A. {Saurous}, B.~{Seybold},
  M.~{Slaney}, R.~J. {Weiss}, and K.~{Wilson},
\newblock ``Cnn architectures for large-scale audio classification,''
\newblock in {\em ICASSP}, 2017.

\bibitem{soundnet}
Y.~Aytar, C.~Vondrick, and A.~Torralba,
\newblock ``Soundnet: Learning sound representations from unlabeled video,''
\newblock in {\em NIPS}. 2016.

\bibitem{Suh2020}
S.~Suh, S.~Park, Y.~Jeong, and T.~Lee,
\newblock ``Designing acoustic scene classification models with {CNN}
  variants,''
\newblock Tech. {R}ep., DCASE2020 Challenge, 2020.

\bibitem{Hu2020}
H.~Hu, C.~H. Yang, X.~Xia, X.~Bai, X.~Tang, Y.~Wang, S.~Niu, L.~Chai, J.~Li,
  H.~Zhu, F.~Bao, Y.~Zhao, S.~M. Siniscalchi, Y.~Wang, J.~Du, and C.~Lee,
\newblock ``Device-robust acoustic scene classification based on two-stage
  categorization and data augmentation,''
\newblock Tech. {R}ep., DCASE2020 Challenge, 2020.

\bibitem{KoutiniEDW19}
K.~Koutini, H.~Eghbal{-}zadeh, M.~Dorfer, and G.~Widmer,
\newblock ``The receptive field as a regularizer in deep convolutional neural
  networks for acoustic scene classification,''
\newblock in {\em EUSIPCO}, 2019.

\bibitem{Chen2019}
H.~Chen, Z.~Liu, Z.~Liu, P.~Zhang, and Y.~Yan,
\newblock ``Integrating the data augmentation scheme with various classifiers
  for acoustic scene modeling,''
\newblock Tech. {R}ep., DCASE2019 Challenge, 2019.

\bibitem{Dorfer2018}
M.~Dorfer, B.~Lehner, H.~Eghbal-zadeh, H.~Christop, P.~Fabian, and W.~Gerhard,
\newblock ``Acoustic scene classification with fully convolutional neural
  networks and {I}-vectors,''
\newblock Tech. {R}ep., DCASE2018 Challenge, 2018.

\bibitem{Liping2018}
Y.~Liping, C.~Xinxing, and T.~Lianjie,
\newblock ``Acoustic scene classification using multi-scale features,''
\newblock Tech. {R}ep., DCASE2018 Challenge, 2018.

\bibitem{pons2017}
J.~{Pons}, O.~{Slizovskaia}, R.~{Gong}, E.~{Gómez}, and X.~{Serra},
\newblock ``Timbre analysis of music audio signals with convolutional neural
  networks,''
\newblock in {\em EUSIPCO}, 2017, pp. 2744--2748.

\bibitem{Huzaifah17}
M.~Huzaifah,
\newblock ``Comparison of time-frequency representations for environmental
  sound classification using convolutional neural networks,''
\newblock {\em CoRR}, vol. abs/1706.07156, 2017.

\bibitem{xiao2020audiovisual}
Fanyi Xiao, Yong~Jae Lee, Kristen Grauman, Jitendra Malik, and Christoph
  Feichtenhofer,
\newblock ``Audiovisual slowfast networks for video recognition,'' 2020.

\bibitem{dap}
Z.~Zhang, Y.~Wang, C.~Gan, J.~Wu, J.~B. Tenenbaum, A.~Torralba, and W.~T.
  Freeman,
\newblock ``Deep audio priors emerge from harmonic convolutional networks,''
\newblock in {\em ICLR}, 2020.

\bibitem{SuZWM19}
Y.~Su, K.~Zhang, J.~Wang, and K.~Madani,
\newblock ``Environment sound classification using a two-stream {CNN} based on
  decision-level fusion,''
\newblock {\em Sensors}, 2019.

\bibitem{Li2019}
X.~Li, V.~Chebiyyam, and K.~Kirchhoff,
\newblock ``{Multi-Stream Network with Temporal Attention for Environmental
  Sound Classification},''
\newblock in {\em Interspeech 2019}, 2019.

\bibitem{Bhatt2018}
G.~Bhatt, A.~Gupta, A.~Arora, and B.~Raman,
\newblock ``{Acoustic features fusion using attentive multi-channel deep
  architecture },''
\newblock in {\em CHiME 2018 Workshop on Speech Processing in Everyday
  Environments}, 2018.

\bibitem{Wang2020_t1}
H.~Wang, D.~Chong, and Y.~Zou,
\newblock ``Acoustic scene classification with multiple decision schemes,''
\newblock Tech. {R}ep., DCASE2020 Challenge, June 2020.

\bibitem{mcdonnell_2020}
M.~{McDonnell} and W.~{Gao},
\newblock ``Acoustic scene classification using deep residual networks with
  late fusion of separated high and low frequency paths,''
\newblock in {\em ICASSP}, 2020.

\bibitem{Wu2020}
Y.~{Wu} and T.~{Lee},
\newblock ``Time-frequency feature decomposition based on sound duration for
  acoustic scene classification,''
\newblock in {\em ICASSP}, 2020.

\bibitem{Han2019}
K.~J. {Han}, R.~{Prieto}, and T.~{Ma},
\newblock ``State-of-the-art speech recognition using multi-stream
  self-attention with dilated 1d convolutions,''
\newblock in {\em ASRU}, 2019, pp. 54--61.

\bibitem{He_2016_CVPR}
K.~He, X.~Zhang, S.~Ren, and J.~Sun,
\newblock ``Deep residual learning for image recognition,''
\newblock in {\em CVPR}, 2016.

\bibitem{Tran_2018_CVPR}
D.~Tran, H.~Wang, L.~Torresani, J.~Ray, Y.~LeCun, and M.~Paluri,
\newblock ``A closer look at spatiotemporal convolutions for action
  recognition,''
\newblock in {\em CVPR}, 2018.

\bibitem{TSN2016ECCV}
L.~Wang, Y.~Xiong, Z.~Wang, Y.~Qiao, D.~Lin, X.~Tang, and L.~{V. Gool},
\newblock ``Temporal segment networks: Towards good practices for deep action
  recognition,''
\newblock in {\em ECCV}, 2016.

\bibitem{Park2019}
D.~S. Park, W.~Chan, Y.~Zhang, C.-C. Chiu, B.~Zoph, E.~D. Cubuk, and Q.~V. Le,
\newblock ``{SpecAugment: A Simple Data Augmentation Method for Automatic
  Speech Recognition},''
\newblock in {\em Interspeech}, 2019.

\bibitem{specaugment_repo}
``Spec{A}ugment,'' https://github.com/zcaceres/spec\_augment.

\end{thebibliography}

\appendix
\section*{Appendix}
In this additional material, we provide further insight into what each of the Slow and Fast streams learn, through class analysis and visualising feature maps from each stream. We also offer an ablation on separable convolutions. Finally, we detail the hyperparameters used to train \cite{mcdonnell_2020} on VGG-Sound. 
\section{Class performance of two streams}
\label{sec:class-level}

\begin{figure*}[t!]
    \centering
    \includegraphics[width=0.5\textwidth]{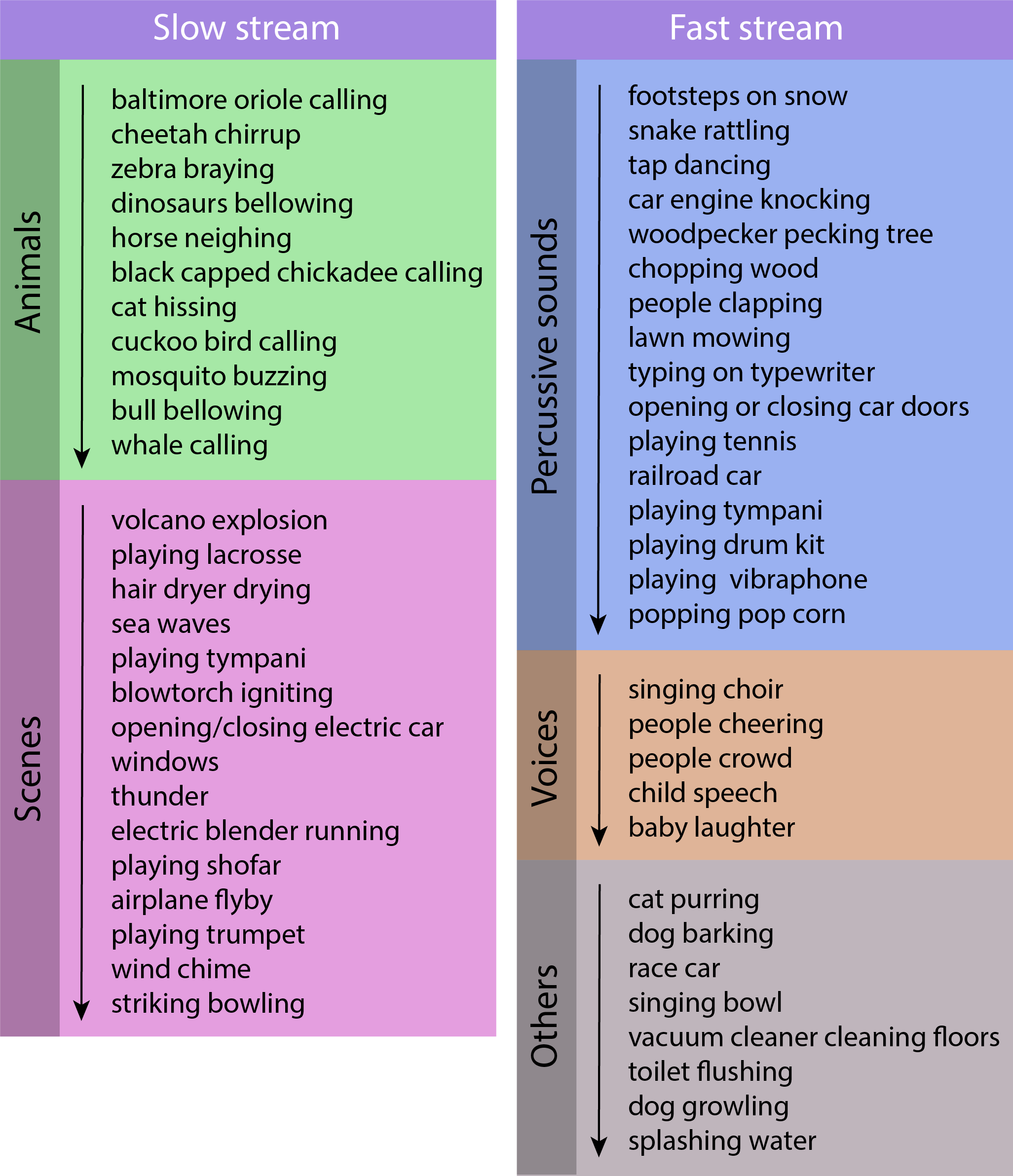}
    \caption{Classes from VGG-Sound that are better predicted from Slow (left) versus Fast (right) streams.}
\label{fig:class_performance}
\end{figure*}

In Figure~\ref{fig:class_performance}, we distinguish between VGG-Sound classes that are better predicted from the Slow stream to the left, and classes that are better predicted from the Fast stream to the right. To obtain these, we calculated per-class accuracy and retrieved classes for which the accuracy difference is above a threshold. Particularly, we used $\text{accuracy}_\text{Slow}-\text{accuracy}_\text{Fast}>20\%$ to retrieve classes best predicted from Slow and $\text{accuracy}_\text{Fast}-\text{accuracy}_\text{Slow}>10\%$ to retrieve classes best predicted from Fast. We used a higher threshold for the Slow stream as it more frequently outperforms the Fast stream, as shown in our earlier results. 

As can be seen in Figure~\ref{fig:class_performance}, Slow predicts better animals and scenes. This matches our intuition that Slow focuses on learning frequency patterns as different animals make distinct sounds at different frequencies, e.g. mosquito buzzing vs whale calling, requiring a network with fine spectral resolution to distinguish between those. In Scenes, there are classes such as sea waves, airplane and wind chime, that contain slow evolving sounds. 

The Fast stream, in contrast, can better predict classes with percussive sounds like playing drum kit, tap dancing, woodpecker pecking tree, and popping popcorn. This also matches our design motivation that the Fast stream learns better temporal patterns as these classes contain temporally localised sounds that require a model with fine temporal resolution.
Interestingly, Fast is better at human speech, laughter, singing, and other human voices, where we speculate that it can better capture articulation.

\section{Visualising Feature Maps}
\label{sec:features}

\begin{figure*}[t!]
    \centering
        \includegraphics[width=\textwidth]{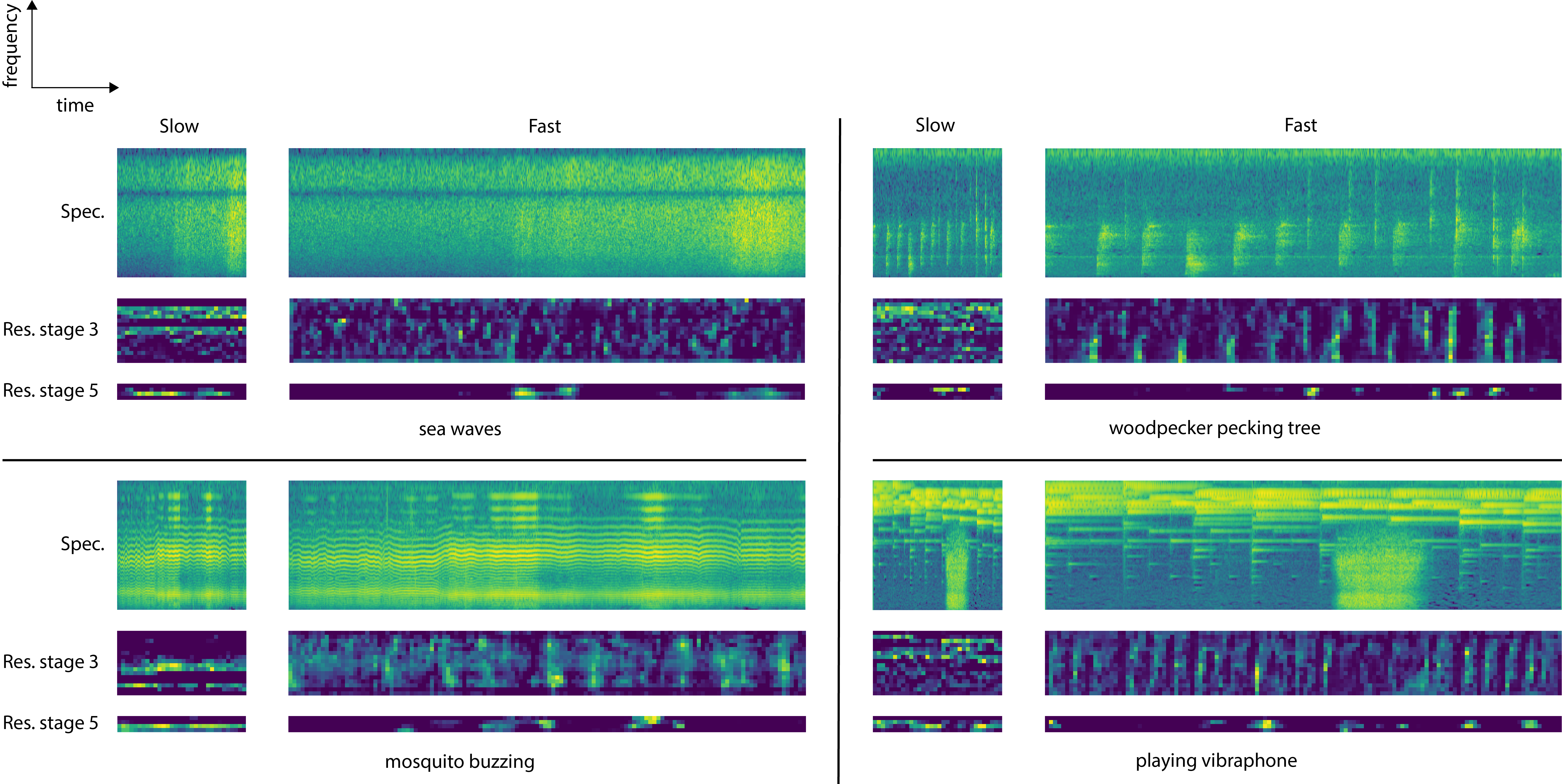}
    \caption{Feature maps from classes that are better predicted from Slow (left) and Fast (right) streams.}

\label{fig:features_fast_better}
\end{figure*}

We show examples of feature maps from Slow and Fast streams, when trained independently (Fig~\ref{fig:features_fast_better}). In each case, we show two samples from classes that are better predicted from the corresponding stream.
For Slow, these are sea waves and mosquito buzzing, compared to woodpecker pecking tree and playing vibraphone for Fast.
In each case, we show the input spectogram as well as feature maps from residual stages 3 and 5.
In each plot, the horizontal axis represents time while the vertical axis corresponds to frequency.
We visualise a single channel from each feature map, manually chosen. 

In Figure~\ref{fig:features_fast_better}, we demonstrate that Fast is capable of detecting the hits of the woodpecker on the tree as well as the hits on the vibraphone, while Slow extracts frequency patterns that do not seem to be useful for discriminating these classes that contain temporally localised sounds. For sea waves and mosquito buzzing, Slow extracts frequency patterns over time, while Fast aims to temporally localise events, which does not assist the discrimination of these classes.

\section{Ablation of separable convolutions}

\begin{table}[t!]
\centering
\resizebox{\linewidth}{!}{%
\begin{tabular}{@{}lrrrrr@{}}
\toprule
                                     Model                                     & \multicolumn{1}{c}{Top-1} & \multicolumn{1}{c}{Top-5} & \multicolumn{1}{c}{mAP}           & \multicolumn{1}{c}{AUC}    & \multicolumn{1}{r}{d-prime}\\ \midrule
                                           Chen et al. \cite{vggsound}     & 51.00 & 76.40 & 0.532 & 0.973 & 2.735\\
                                          ResNet50 & 52.23 & 78.08 & 0.542 & 0.974 & 2.747 \\
                                          ResNet50-separable   & 52.38 & 77.81 & \textbf{0.544} & \textbf{0.975} & \textbf{2.777}\\
            							 Slow-Fast (Proposed)     & \textbf{52.46} & \textbf{78.12} & \textbf{0.544} & 0.974 & 2.761\\
                                          \bottomrule
\end{tabular}}
\caption{Ablation of separable convolutions on VGG-Sound.}
\label{tab:separable}
\end{table}

We provide an ablation of separable convolutions in Table~\ref{tab:separable}. We trained the ResNet50 architecture as proposed in \cite{He_2016_CVPR} without separable convolutions, as well as a variant with separable convolutions. We compare this to the published results by Chen et al. \cite{vggsound}  that also uses a ResNet50 architecture.
Our reproduced results already outperform~\cite{vggsound}.
ResNet50-separable has separable convolutions as used in our Slow-Fast network (see Figure~\ref{fig:arch} and Table~\ref{tab:arch}). 

Results show that ResNet50-separable achieves slightly better results than ResNet50 in all metrics except Top-5.
Although accuracy is not significantly increased in this ablation, we employ separable convolutions in our proposed architecture, following our motivation to attend differently to frequency and time.
These results also show that a single stream ResNet50 has comparable performance to our two stream proposal, however ours performs better in accuracy and the two streams accommodate different characteristics of audio classes as shown previously. 

\section{Hyperparameter Details}
Training the publicly available code of McDonnell \& Gao \cite{mcdonnell_2020} with the default hyperparameters on VGG-Sound provided poor results. We tuned the hyperparameters as follows: We set the maximum learning rate to $0.01$, train the network for $62$ epochs, with $\text{alpha}=0.1$ for mixup. Lastly, we adjusted the number of FFT points to $682$ for log-mel-spectrogram extraction, to apply a window and hop length similar to the ones in \cite{mcdonnell_2020} (their datasets are sampled at 48kHz and 44.1kHz, while VGG-Sound is sampled at 16kHz).
\end{document}